\documentclass[sigconf]{acmart}

\usepackage{times}
\usepackage{paralist}
\usepackage{enumitem}
\usepackage{textcomp}
\usepackage{bibentry}
\usepackage{color}
\usepackage{graphicx}
\usepackage{url}
\usepackage{xspace}
\usepackage{multirow}

\usepackage[compact]{titlesec}
\titlespacing*{\section}{2pt}{*1}{*0}
\titlespacing*{\subsection}{1pt}{*0.5}{*0}
\titlespacing*{\subsubsection}{1pt}{*0}{*1}

\setcopyright{acmcopyright}
\copyrightyear{2018}
\acmYear{2020}
\acmDOI{10.1145/1122445.1122456}

\acmConference[Supercomputing '20]{Supercomputing '20: International Conference for High Performance Computing, Networking, Storage, and Analysis}{November 16--19, 2020}{Virtual}
\acmBooktitle{Supercomputing '20: International Conference for High Performance Computing, Networking, Storage, and Analysis}

\begin{document}

\newif\ifdraft

\ifdraft
 \newcommand{\jhanote}[1]{{\textcolor{red}{ ***Shantenu: #1 }}\xspace}
 \newcommand{\note}[1]{{\textcolor{blue}{ ***Note: #1 }}\xspace}
\else
 \newcommand{\jhanote}[1]{}
 \newcommand{\note}[1]{}
\fi

\newcommand{\project}{IMPECCABLE }

\title{IMPECCABLE: Integrated Modeling PipelinE for COVID Cure by Assessing Better LEads}

\author{Aymen Al Saadi$^1$, Dario Alfe$^{2,10}$, Yadu Babuji$^3$, Agastya Bhati$^2$, Ben Blaiszik$^{3,4}$, Thomas Brettin$^4$, Kyle Chard$^{3,4}$, Ryan Chard$^3$, Peter Coveney$^{*2,9}$, Anda Trifan$^4$,
Alex Brace$^4$, Austin Clyde$^2$, Ian Foster$^{3,4}$, Tom Gibbs$^8$, Shantenu
Jha$^{*1,5}$, Kristopher Keipert$^8$, Thorsten Kurth$^8$, Dieter
Kranzlmüller$^7$, Hyungro Lee$^1$, Zhuozhao Li$^3$, Heng Ma$^4$, Andre Merzky$^1$, Gerald
Mathias$^7$, Alexander Partin$^4$, Junqi Yin$^6$, Arvind Ramanathan$^{*4}$, Ashka Shah$^4$, Abraham Stern$^8$, Rick
Stevens$^{*3,4}$, Li Tan$^5$, Mikhail Titov$^1$, Aristeidis Tsaris$^6$, Matteo
Turilli$^1$, Huub Van Dam$^5$, Shunzhou Wan$^2$, David Wifling$^7$}
\affiliation{$^1$ Rutgers University, $^2$ University College London, $^3$ University of Chicago, $^4$ Argonne National Laboratory, \\ $^5$ Brookhaven National Laboratory,
$^6$ Oak Ridge Leadership Computing Facility, $^7$ Leibniz Supercomputing
Centre, \\ $^8$ NVIDIA Corporation, $^9$University of Amsterdam, $^{10}$University of Naples Federico II, $^*$ Contact Authors}

\renewcommand{\shortauthors}{Al Saadi, Alfe, et al.}

\begin{abstract}
The drug discovery process currently employed in the pharmaceutical industry
typically requires about 10 years and \$2-3 billion to deliver one new drug.
This is both too expensive and too slow, especially in emergencies like the COVID-19
pandemic. In silico methodologies need to be improved to better select lead
compounds that can proceed to later stages of the drug discovery protocol accelerating the entire process.  No single methodological approach can achieve
the necessary accuracy with required efficiency. Here we describe multiple
algorithmic innovations to overcome this fundamental limitation, development
and deployment of computational infrastructure at scale integrates multiple artificial intelligence
and simulation-based approaches. Three measures of performance are:  (i)
throughput, the number of ligands per unit time;
(ii) scientific performance, the number of
effective ligands sampled per unit time; and (iii) peak performance,
in flop/s. The capabilities outlined here have been used in
production for several months as the workhorse of the the computational
infrastructure to support the capabilities of the US-DOE National Virtual
Biotechnology Laboratory in combination with resources from the EU Centre of Excellence in Computational Biomedicine.

\end{abstract}

\keywords{Datasets, neural networks, gaze detection, text tagging, docking molecular dynamics, free energy estimation}

\maketitle

\section{Justification}

COVID-19 has claimed a million lives and resulted in over 35 million
infections; there is an urgent need to identify drugs that can inhibit
SARS-CoV-2. \project innovatively couples multiple algorithms to overcome
fundamental limitations of classical \textit{in silico} drug design. We
discuss how algorithmic and computational innovations are advancing scientific
discovery.

\section{Performance Attributes}

Performance attributes are listed in Table~\ref{tab:perf-attr}.

\begin{table}[h!]
\centering
\caption{Performance Attributes}\label{tab:perf-attr}

\vspace{-3mm}

\begin{tabular}{ll}
\hline
Performance Attribute & Our Submission  \\ 
\hline
\hline
Category of achievement & Scalability, time-to-solution, \\
                        & peak performance \\ 
Type of method used & Explicit \\
Results reported on basis of &  Whole app including I/O \\
Precision reported & Mixed  \\
System scale & Measured on full system  \\
Measurement mechanism & Timers, FLOP count, \\ 
& performance modeling\\
\hline            
\end{tabular}
\end{table}

\section{Problem Overview}

Drug discovery is an astonishingly resource intensive process; the
average time to search, design, and effectively bring a clinically tested drug can range between 10 to 15 years, and can cost over 1~billion dollars~\cite{Hughes_et_al:2010,saha2020pharmaceutical}.
Considering the universe of about 10$^{68}$ possible compounds to traverse for
effective drugs, there is an immediate need for more efficient, higher
throughput, and more meaningful frameworks for early stage drug discovery~\cite{Bohacek_et_al:2010}.

In the context of COVID-19, a grand challenge within the drug-discovery
community is the need for capabilities that can screen {\it tens of billions}
of small molecules against the SARS-CoV-2 proteome and identify high quality
lead molecules that have the potential to inhibit the virus life
cycle~\cite{Parks_2020}. To achieve this goal, \textit{in silico}
methodologies need to be significantly improved to better design lead
compounds that have the potential to become drugs that can be pushed to the
later stages of drug
discovery~\cite{antunes2015understanding,zhou2020artificial,smith2018transforming}.
However, no single algorithm or method can achieve the necessary accuracy with
required efficiency. \project innovatively brings together multiple algorithms
into a single unified pipeline with an interactive and iterative methodology
allowing both upstream and downstream feedback to overcome fundamental
limitations of classical \textit{in silico} drug design. We demonstrate the
impact of \project by measuring both raw throughput,
defined as ligands per unit time, as well as its
scientific performance, defined as ligands
sampled per unit time.

\subsection{Targeting the SARS-CoV-2 proteome} 

The SARS-CoV-2 genome consists of 29 proteins of which 16 non-structural
proteins (NSPs) represent various enzymes that play critical roles in the
virus life cycle~\cite{Parks_2020}. In spite of our understanding of the roles of these NSPs in
the virus life cycle as well as the fact that closely related protein
structures from the middle eastern respiratory syndrome (MERS) and the SARS
coronaviruses are known, \emph{there are no known antivirals currently available for SARS-CoV-2}~\cite{Wang_2020}. The structural biology community has
embarked on a massive effort to provide access to 3D crystallographic
structures of the NSPs and therefore, this represents a tremendous opportunity
for designing therapeutics targeting the disease.

Early stages of drug discovery rely on (experimental) high throughput
screening (HTS) protocols to hone in on a suitable chemical compound library
that can be particularly useful against known protein
targets~\cite{PMID:8895594}. Although HTS approaches are widely available and
used, the sheer combinatorics of drug-like molecules poses a tremendous
challenge in exhaustively sampling the compound space to find viable
molecules. Instead, a number of open source initiatives are currently
building virtual HTS platforms targeting the entire viral
proteome~\cite{moonshot,molssi}.
Indeed, the goals of our efforts include high throughput structure-based
protein-ligand docking simulations, followed by iterative refinements to these
virtual screening results to filter out compounds that ``show promise'' in
biochemical or whole-cell assays, safety and toxicology tests, finally leading
to a set of compounds that can proceed towards clinical trials. This process
represents a filtering process, with successive steps leading to better drug
candidates targeting the virus. Although covering such a vast chemical space
would be nearly impossible, even with access to the entire world's computing
resources, there are open data resources which provide
access to a fairly representative, yet diverse chemical space ($\sim$10$^{12}$
compounds).

\subsection{The IMPECCABLE Solution} \label{sec:ImpeccableSoln}
Artificial intelligence (AI) and machine learning (ML) have played
a pivotal role in COVID-19 drug discovery~\cite{zhou2020artificial}. However, most ML/AI efforts have
largely focused on building effective means to analyze large volumes of
data generated through either ligand docking simulations---for the
purposes of filtering favorable vs.\ unfavorable ligand binding poses in a
given protein---or 
molecular dynamics (MD)
simulations of selected protein-ligand complexes.

While docking programs are generally good at pose prediction, they are less
effective in predicting binding free-energy of protein-ligand complexes.
Similarly, while MD simulations are effective at predicting binding-free
energies, their intrinsic limitations in sampling protein-ligand complex
formation processes imply that the approach may be computationally infeasible to
translate on large compound libraries. If ML/AI methods can \emph{glue}
information across docking and MD simulation techniques, while leveraging
their individual strengths to provide meaningful feedback in terms of
identifying parts of the compound libraries that may result in better lead
molecules, higher effective throughput (sampling larger compound libraries),
and meaningful enrichment of lead molecules for targeting 
SARS-CoV-2 proteins would be possible.

\begin{figure}[h]
  \centering
  \includegraphics[width=\linewidth]{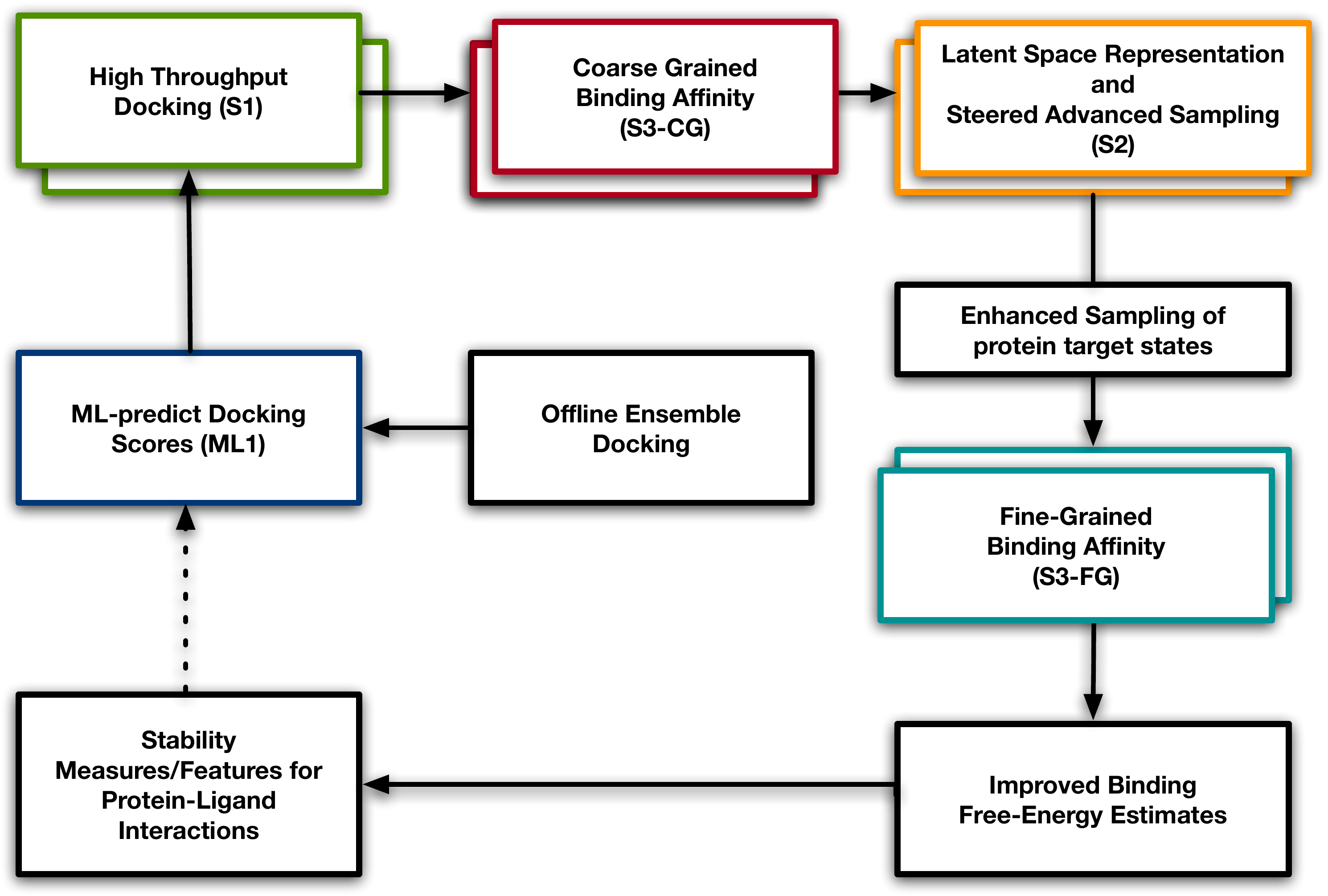}
  
    \vspace{-3mm}
    
  \caption{The IMPECCABLE Solution:  represents an entire virtual drug discovery pipeline, from hit to lead
  through to lead optimization. The constituent components are deep-learning
  based surrogate model for docking (ML1), Autodock-GPU (S1), coarse and
  fine-grained binding free energies (S3-CG and S3-FG) and S2
  (DeepDriveMD).}\label{fig:impeccablesoln}
  \Description{Integrated and multistage Campaign}
 \vspace{-0.1in}
\end{figure}

\project addresses these scientific and
methodological challenges by providing a scalable, flexible, and extensible
infrastructure for campaigns designed to discover improved leads targeted at
SARS-CoV-2. As shown in Fig. \ref{fig:impeccablesoln}, the \project campaign
consists of an iterative loop initiated with ML predictions (ML1), followed by three stages of data processing (S1, S2, S3). \project centers on the use of ML/AI
techniques (ML1 and S2) interfaced with physics-based computational
methods to estimate docking poses of compounds that are promising leads
for a given protein target (S1) and binding free-energy computations (S3).

ML techniques overcome the limitations of S1 and S3 by predicting the
likelihood of binding between small molecules and a protein target (ML1), and
accelerating the sampling of conformational landscapes to bound the binding
free-energy values for a given protein-ligand complex (S2). Interfacing ML
approaches with physics-based models (docking and MD simulations), we
achieve at least three orders of magnitude improvement in the size
of compound libraries that can be screened with traditional approaches, while
simultaneously providing access to binding free-energy calculations that can
impose better confidence intervals in the ligands selected for further 
(experimental or computational) optimization.

\subsubsection*{ML1: Machine Learning Models for docking score prediction}

Scoring functions are used to score poses in order to determine the most
likely pose of the molecule, the magnitude of which is used to provide an indication
of active versus inactive ligands, and lastly to rank order sets of libraries.
We create a ML surrogate model to replace the use of docking as a means of
locating regions of chemical space likely to include strong binding drug leads.
The only free variable for the surrogate ML ranking function is the basic
molecular information, which typically presents as a SMILES string. We use a
simple featurization method, namely 2D image depictions, as they do not require
complicated architectures such as graph convolution networks, while
demonstrating good prediction.
We obtain these image depictions from the nCov-Group Data Repository~\cite{babuji2020targeting}, 
which contains various descriptors for 4.2B molecules generated on HPC systems with the assistance of Parsl~\cite{babuji2019parsl}.

\subsubsection*{S1: High-throughput Docking}

Protein-ligand docking encompasses a computational pipeline consisting of
ligand 3D structure (conformer) enumeration, exhaustive docking and scoring,
and final pose scoring. The input to the docking protocol requires a protein
structure with a designed binding region, or a crystallized ligand from which
a region can be inferred, as well as a database of molecules to dock in SMILES
format. SMILES format is a compact representation of a 2D molecule.
     
\subsubsection*{S2: Machine Learning Driven Molecular Dynamics}

Machine learning tools are able to quantify statistical insights into the
time-dependent structural changes a biomolecule undergoes in simulations,
identify events that characterize large-scale conformational changes at
multiple timescales, build low-dimensional representations of simulation data
capturing biophysical/biochemical/biological information, use these
low-dimensional representations to infer kinetically and energetically
coherent conformational substates, and obtain quantitative comparisons with
experiments.

Deep structured learning approaches automatically learn lower-level
representations (or features) from the input data and successively aggregating
them such that they can be used in a variety of supervised, semi-supervised
and unsupervised machine learning tasks. We developed variational
autoencoders to automatically reduce the high dimensionality of MD
trajectories and cluster conformations into a small number of conformational
states that share similar structural, and energetic
characteristics~\cite{Bhowmik_2018}. We use S2 to drive adaptive sampling
simulations, and use the acceleration of ``rare'' events, to investigate
protein-ligand interactions~\cite{Lee_2019,Romero_2019}.

\subsubsection*{S3: Binding Free Energy Calculations}

Hit-to-Lead (H2L), sometimes also called lead generation, is a step in the
drug discovery process where promising lead compounds are identified from
initial hits generated at preceding stages. It involves evaluation of initial
hits followed by some optimization of potentially good compounds to achieve
nanomolar affinities. The change in free energy between free and bound states
of protein and ligand, also known as binding affinity, is a promising measure
of the binding potency of a molecule, and hence it is used as a parameter for
evaluating and optimizing hits at H2L stage. We employ the ESMACS protocol
\cite{brd4,trka,esmacs1,esmacs2,UCL_review}, for estimating binding affinities
of protein-ligand complexes. We differentiate between
coarse-grained (CG) and fine-grained (FG) ESMACS variants, which differ in
the number of replicas (6 vs.\ 24), equilibration duration (1 vs.\ 2ns),
simulation duration (4 vs.\ 10ns) etc. The computational cost of ESMACS-CG is
about an order of magnitude less than that of ESMACS-FG.

\project integrates multiple methods and dynamically selects active ligands
for progressively expensive methods. In fact, at any stage, only the most
promising candidates are advanced to the next stage, yielding a N-deep
pipeline, where each downstream stage is computationally more expensive, but
also more accurate than previous stages. The methods chosen vary in
computational cost per ligand by more than six orders of magnitude; in the
docking stage of \project each dock costs about $10^{-4}$ node-hours per
ligand; fine-grained binding free energy costs about $10^2$ node-hours per
ligand. This provides an important dynamic range of accuracy, and thus
potential for scientific performance enhancement. Tuning the cost of each
method by extending or contracting the number of iterations of each method
allows for enhanced scientific performance and throughput.

The integration of methods with varying computational characteristics into a
cohesive whole to support sustained computational campaign requires innovative
computational infrastructure. There are no turnkey or shrink-wrap solutions to
support campaigns. We employ RADICAL-Cybertools (RCT)~\cite{github-rct}, a set of software systems, to manage the execution of
heterogeneous workflows and workloads on leadership computing facilities. RCT
have the required design and necessary abstractions to manage the
heterogeneity of workloads and platforms, and the integrated campaign.

\section{State of the Art}

We outline briefly the current state-of-the-art with respect to our
computational campaign, summarizing some of the challenges and limitations
in the context of \project.

The idea of building large-scale virtual screening pipelines is not entirely new. However, many of these toolkits are either proprietary (e.g., Schrodinger, OpenEye) or even if they are openly available (e.g., AMPL~\cite{minnich2020ampl}), they end up being very customized. Some of the early approaches~\cite{Zhang_2014,Spyrakis_2015,Kumar_2015} proposed the use of statistical techniques to be incorporated -- either for downstream refinement or for selection of viable leads -- however, sheer size of ligand docking simulations as well as MD simulations could easily overwhelm these approaches, leading to very little gain from the overall process. With the advent of deep learning~\cite{Bengio_2015} and recent advances in computing technology for running MD simulations, the burden of handling large volumes of virtual screening data and simulations has been largely reduced. This in turn has spurred momentum in developing several automated pipelines whose goal is to improve not just the virtual screening process itself, but also to enable downstream processes such as lead refinement and optimization. 

Central to most high throughput virtual screening pipelines are scoring functions that are used to determine
which pose is selected via exhaustive search and how the resulting
molecule and pose rank within a dataset. While some docking
protocols expand the typical exhaustive mapping of a scoring function over
possible positions, the distinction between protocols boils down to the
scoring function. 

The scoring function is used for three purposes:
(i) to detect when the ligand is properly positioned in the pocket (\textit{pose prediction}); (ii) to provide a general assessment of activity (\textit{virtual screening}), and (iii) to rank ordering compound libraries from their selected poses (\textit{binding affinity ranking}) \cite{guedes2018empirical, clyde2020regression}. Given the central role of scoring functions, the enterprise of
molecular docking pragmatically and theoretically rests on the speed and
accuracy of these scoring functions. Some scoring functions can be slow, and
particularly well suited for pose prediction, while others may be fast and only
capable of decoy detection. However, scoring functions by themselves are not sufficient; they need to be augmented with physically accessible (and experimentally verifiable) quantities such as binding free-energies (or binding affinities). 

There are a large number of \textit{in silico} methods available to calculate
binding affinities. The major ones, in the order of increasing accuracy as
well as computational expense, are as follows: (1) molecular
docking~\cite{docking2,docking3}, (2) linear interaction energy~\cite{LIE}, (3)
MMPBSA/MMGBSA and (4) alchemical methods (Thermodynamic Integration/Free
Energy Perturbation)~\cite{ti1,fep}. Alchemical methods are theoretically the
most exact whereas the other three involve approximations to some extent,
decreasing in the above order. This is why we need to choose an appropriate
method for each step in the entire drug discovery process, keeping in mind the
level of accuracy desired and the number of compounds to be considered. In our
workflow, we employ the cheapest and not greatly accurate docking methods at
the initial stages, MMPBSA-based ESMACS at the hit-to-lead stage and TI based
TIES at the lead optimization stage. The throughput reduces by orders of
magnitude as we move to subsequent stages with higher levels of accuracy
desired and hence the computational cost remains under control.

As demonstrated in Tab.~\ref{tab:cost}, \project integrates methods
with 6--7 orders of magnitude computational cost, and although hitherto not part of our
demonstrated scientific impact, can support TIES which is a further two orders
of magnitude more expensive than ESMACS. Thus, \project represents a unique
solution that integrates multiple methods, with a collective
and integrated performance significantly greater than any single
algorithm and method alone.

\section{Innovations Realized}

\project embodies innovation within the individual methods it employs, as well as
in the way it couples these methods. Underpinning this coupling is a flexible
and scalable infrastructure. The ability to potentially screen much larger libraries with higher
throughputs so as to identify greater number of of viable SARS-CoV-2 protein
target specific leads is the ultimate measure of scientific productivity. To
that end, we characterize the contribution of each component, as well as the
improvement in performance of the IMPECCABLE pipeline by integrating
individual components. Actual results are presented in
Section~\ref{sec:results}.

\subsection{Algorithmic \& Methodological}

\subsubsection{S1: AutoDock-GPU} 

The CUDA-accelerated
AutoDock4.2.6 (AutoDockGPU) leverages a highly
parallel implementation of the Lamarckian genetic algorithm (LGA) by processing ligand-receptor poses in parallel over
multiple compute units. AutodockGPU was developed in collaboration with NVIDIA
and Aaron Scheinberg (Jubilee Development) with a target of the Summit system
at Oak Ridge Leadership Computing Facility (OLCF). AutoDockGPU applies the
legacy Solis-Wets local search method along with a new local-search method
based on gradients of the scoring function. One of these methods, ADADELTA,
has proven to increase significantly the docking quality in terms of RMSDs and
scores with observed speedups of 56x over the original serial AutoDock 4.2
(Solis-Wets) on CPU. A drug screen takes the best scoring pose from these
independent outputs. Autodock-GPU uses OpenMP threading-based pipeline for
hiding ligand input and staging, and the receptor-reuse functionality for
docking many ligands to a single receptor. From the
computational performance perspective, we measure the total number of docking
calculations performed per GPU, which provides a measure of the overall
docking capability.

\subsubsection{ML1: ML-based docking score predictor}
We developed a ML surrogate model to replace docking as means of
locating regions of chemical space likely to include strong binding drug
leads. Docking does not rely on proteins, instead proteins are coded into the
scoring function. Thus the only free variable for the surrogate ML ranking
function is the basic molecular information. This usually presents as
SMILES string, and there is an entire field of deep learning for molecular
property prediction based on this approach~\cite{Elton_2019}.

A simple featurization method has been widely ignored---2D image
depictions. From the 2D depiction of a molecule, chemists are generally able
to identify major properties such as H-acceptors, estimate the molecule's
weight, and even determine if a molecule might bind to a protein~\cite{goh2017chemception,MolNet_2018,ramsundar2015massively}. This
featurization method, unlike graph structure, is able to utilize off-the-shelf
convolutional neural networks. By using 2D images, we are able to initialize
our models with pretrained weights that are typically scale and rotation
invariant for image classification tasks, which is exactly we require in order to infer if a small molecule will bind well to a given SARS-CoV-2 target.

We model the
gains and losses using a cost function based on the regression
enrichment surface (RES)\cite{clyde2020regression}. The RES measures how well
a surrogate model can detect the true top ranking molecules given a certain
allocation of predicted hits. 

While this analysis brings forth the failure of the model to exactly replicate
the rank ordering of the compounds at scale, it provides the operational
benefit of these models---the predictive ML model will indeed be able to
filter with near 100\% accuracy two orders of magnitude from the data library.
Thus, if all else is equal, with only the additional cost of
docking additional compounds, we are able to expand the set of viable leads detected by two orders of
magnitude, without loss of performance in the top regions of detection (Sec. \ref{sec:results}). 

\subsubsection{S3: Adaptive ESMACS}
Binding affinity is a small number (a few tens of kcal/mol) that is derived
from absolute free energies which are large (a few hundreds to thousands of
kcal/mol). Thus, the usual practice of performing MMPBSA calculations on
conformations generated using a single MD simulation does not give reliable
binding affinities. ESMACS, on the other hand, performs ensemble MD
simulation, where each independent simulation is termed a \emph{replica}. 
Parameters such as the size of ensemble simulation (or the number of replicas)
and the length of individual replica are chosen such that our results become
reliable quantities~\cite{hiv1, hiv2}. Another factor that plays a role
in determination of these parameters is the level of precision desired and the
cost-benefit ratio. The number of replicas performed is adjusted to a find a
sweet spot between computational cost and the level of precision acceptable at
a particular stage of the pipeline.

ESMACS  is costlier than the
standard approach of performing a single simulation of
similar duration.
This increased cost however, is more than compensated by the enhanced precision
of ESMACS results which makes the resultant ranking of compounds much more
reliable compared to standard approaches with similar accuracy. MMPBSA based
free energies have huge variability in results rendering them
non-reproducible~\cite{hiv1, hiv2, pmhc, brd4, trka}. In fact, fewer
iterations are required to achieve the same level of convergence in
chemical space on using ensemble simulation based methods, which leads to
comparable (or even reduced) computational cost overall, than on using standard single
simulation approaches. This apparent increased cost has advantages, such as an
increased level of confidence in predicted ranking of compounds, and thus much
more reliable training data for an ML model.

We used ESMACS-CG to perform the initial screening of thousands of hits in order
to reduce computational cost while compromising on the level
of precision and ranking of compounds, and used ESMACS-FG for
the latter stages when we have better binding poses, and/or LPC conformations. Selectively using ESMACS-FG on a refined set
of complexes decreases the computational cost substantially without affecting
the quality of results.

\subsubsection{S2: AI-Driven MD}\label{sec:AIDMD}  We leverage 
DeepDriveMD~\cite{lee2019deepdrivemd} to simulate large ensembles of
protein-ligand complexes. We have shown that DeepDriveMD can potentially
accelerate protein folding simulations by at least 2 orders of magnitude. Here
DeepDriveMD (S2) builds an adaptive sampling framework to support the
exploration of protein-ligand bound states that are not often accessible to
approaches such as ESMACS (S3).

A key innovation is support for extremely large numbers of ligand-protein
complexes (LPC). This stems from the fact that a ESMACS-CG (S3-CG) simulations
may generate on average, six ensembles which are analyzed by our novel MD-driven
AI approaches to identify 5--10 novel states. Hence, we also
implemented a novel approach for analyzing large MD ensemble simulation
datasets using a 3D adversarial autoencoder (3D-AAE), a
significant improvement over approaches such as variational autoencoders in
that it is more robust and generalizable to protein coordinate datasets than
contact maps (or other raw inputs) extracted from MD simulations.
Similarly to autoencoders, 3D-AAE builds a latent embedding space for MD
simulations to characterize conformational changes within protein-ligand
complexes from ESMACS-CG/FG simulation trajectories. The 3D-AAE includes the
PointNet encoder, Chamfer distance-based reconstruction loss, and a Wasserstein
adversarial loss with gradient penalty to build a latent manifold on which
all simulations are projected. From this latent manifold, we use local outlier
factor (LOF) detection to identify `interesting' protein-ligand complexes that
are then selected for S3-FG simulations. The iterative feedback between two
stages of S3-CG/FG and S2 enables accurate estimates for the
binding free-energy, and allows us to filter compounds based on their
affinity to the protein, while accounting for the intrinsic conformational
flexibility of the LPC.

Measuring DeepDriveMD performance for LPCs presents
additional challenges. For example, input from the S3-CG pipeline stage are
relatively short time-scale, whereas LPC association processes tend to vary
significantly in time scales. Thus, we chose a pragmatic measure of
LPC stability that takes into account the number of heavy atom
contacts between the protein and the ligand of interest. From the top scoring
LPCs that are selected from S3-CG, we use a novel 3D-AAE to \emph{filter}
those conformations that show increased stability profiles in the LPCs. We
posit that these LPCs are of the most interest, since the increased stability
potentially contributes to favorable interactions between protein and
ligand. We also measure 3D-AAE performance in terms of its ability
to \emph{learn} effective latent space representations from S3-CG stage
(through standard measures such as training and validation loss metrics).

\begin{table}[h]
  \caption{Normalized computational costs on Summit.}\label{tab:cost}
  
  \vspace{-3mm}

 \centering
  \begin{tabular}{l|rrr}
   \hline
    Method&
    \vtop{\hbox{\strut Nodes per}\hbox{\strut \hspace{2 mm} ligand}}&
    \vtop{\hbox{\strut Hours per}\hbox{\strut \hspace{1 mm} ligand}\hbox{\strut (approx)}}&
    \vtop{\hbox{\strut Node-hours}\hbox{\strut \hspace{0.5 mm} per ligand}}\\
    \hline
    \hline
    Docking (S1)           & 1/6 & 0.0001 & $\sim$0.0001 \\
    BFE-CG  (S3-CG)           & 1 & 0.5 & 0.5 \\
    Ad. Sampling (S2) & 2 & 2 & 4  \\
    BFE-FG   (S3-FG)         & 4 & 1.25 & 5 \\
    BFE-TI (not integrated) & 64 & 10 & 640 \\
    \hline
   \end{tabular}
   \vspace{-5mm}
\end{table}

\paragraph{Putting it together and collective performance.}

Each stage of \project when augmented with relatively simple ML/AI approaches
provides a significant boost to the coverage of the compound diversity as well
as conformational landscapes of protein-ligand complexes. Using training data
generated on small O($10^6$) compound libraries, ML1 enables a significant
improvement in filtering large compound O($10^9$) libraries, increasing the
coverage by 4--6 orders of magnitude.

The second step, which results in filtering the top 1\% of these compounds
(which can also set by the end-user) through AutoDock-GPU, identifies high
confidence lead molecules that can bind to a given SARS-CoV-2 target. The
purpose of the ML1 is to predict if the given molecule will dock the
protein well, and not to predict the docking pose. We exploit the intrinsic
strengths of most docking programs in predicting the binding pose for a given
LPC, such that the intial poses selected follow physical principles (i.e.,
optimizing electrostatic and hydrophobic complementarity).

The next stage, namely S3-CG refines the filtered compounds to obtain an
estimate of the binding-free energy. This step is crucial in the sense that it
seeds the further pipeline with higher confidence leads that may have
favorable interactions with the protein target.

This set of diverse LPC are input to S2, which leverages the 3D-AAE to learn a
latent manifold that consists of a description of which LPCs are most stable.
In addition, the latent manifold also captures intrinsic dimensions of the
protein's conformational landscape that are perturbed by the ligand's
interactions. Using outlier detection methods, we then filter these landscapes
further to include only a small number of LPCs on which SG-FG  are implemented
to ultimately suggest strong confidence intervals for binding free-energy of
the LPCs selected.

The final stage of the pipeline provides additional features that identify key
complementarity features (e.g., electrostatic interactions through hydrogen
bonds or hydrophobic interactions) that can be input to the docking program
for further refinement. Each successive iteration of \project thus provides
successive yields of LPCs that could be modeled as an active learning pipeline
for obtaining highly specific small-molecules that can inhibit a SARS-CoV-2
protein of interest.

\subsection{Computational Infrastructure}

The integration of aforementioned diverse methods with varying computational
characteristics, performance and scalability challenges, into a dynamic and
adaptive computational campaign  requires innovative computational
infrastructure. The campaign workload is a diverse mix of task types, e.g.,
MPI, single GPU, multinode GPU and regular CPU; this mix of tasks changes
over the course of the campaign There are multiple stages which couple and
concurrently execute deep learning and traditional simulations. Coupling and
concurrently executing these diverse tasks is challenging, but is made more
difficult by virtue of having different models and coupling with simulations
across multiple stages.

The dynamic variation of workload arises due to many reasons, for example: (i)
adaptive methods, e.g., each LPC has a different rate of convergence for
structural and energetic properties, and thus the duration varies; (ii)
cost of docking per ligand varies across different drug compound libraries and
the ligands they contain; and (iii) for methods that involve learning,
(re-)~training times are dependent on specific ligands and the number of
simulations.

\project employs the Ensemble Toolkit
(EnTK)\cite{balasubramanian2018harnessing}, which uses RADICAL-Pilot
(RP)~\cite{merzky2018using} for flexible and scalable execution of workflows
with heterogeneous tasks. Together they conform to the middleware building
blocks architectural pattern \textemdash{} which constitute recent advances in
the science of HPC workflows~\cite{turilli2019middleware} \textemdash{} to
permit a decoupling of the programming system from underlying execution
capabilities.

\subsubsection{Programming System} EnTK
is a Python implementation of a workflow engine, designed to support the
programming and execution of applications with ensembles of tasks. EnTK
executes tasks concurrently or sequentially, depending on their arbitrary
priority relation. We use the term ``task'' to indicate a stand-alone process
that has well-defined input, output, termination criteria, and dedicated
resources. For example, a task can indicate an executable which performs a
simulation or a data processing analysis, executing on one or more nodes on
Summit. Tasks are grouped into stages and stages into pipelines depending on
the priority relation among tasks. Tasks without a reciprocal priority relation
can be grouped into the same stage, whereas tasks that need to be executed before
other tasks have to be grouped into different stages. Stages are then grouped
into pipelines and, in turn, multiple pipelines can be executed either
concurrently or sequentially. Specifically, EnTK:

\begin{asparaitem}	
\item permits asynchronous execution of concurrent pipelines (each pipeline can
    progress at its own pace);
    \item allows arbitrary sizing of stages (variable concurrency);
	\item supports heterogeneous tasks of arbitrary types, and combinations,
	as well as their inter-mixing;
	\item promotes ``ensembles'' as first-class code abstraction;
	\item selects parameters at runtime so as to provide near-optimal selection of
cost versus accuracy~\cite{scale18dakka,escience18dakka}.

\end{asparaitem}

These are necessary capabilities to explore LPCs of varying
complexity and cost, without constraining the number of concurrent
investigations, and different methods run in arbitrary order.

\subsubsection{Execution Framework for Dynamic Resource Management}

Given the extreme workload heterogeneity and workload variation between and
across stages, dynamic resource management is critical. Dynamic resource
management capability is provided by
RADICAL-Pilot (RP)~\cite{merzky2018using,turilli2019characterizing}, a Python
implementation of the pilot paradigm and architectural
pattern~\cite{turilli2018comprehensive}. Pilot systems enable users to submit
pilot jobs to computing infrastructures and then use the resources acquired by
the pilot to execute one or more workloads, i.e., set of tasks. Tasks are
executed concurrently and sequentially, depending on the available resources.
For example, given 10,000 single-node tasks and 1000 nodes, a pilot system will
execute 1000 tasks concurrently and each one on the remaining 9000 tasks
sequentially, whenever a node becomes available. RP enables the execution of
heterogeneous workloads comprised of one or more scalar, MPI, OpenMP,
multi-process, and multi-threaded tasks. RP directly schedules and executes on
the resources of one or more pilots without having to use the infrastructure's
batch system. 

RP offers unique features when compared to other pilot systems or tools that
enable the execution of multi-task workloads on HPC systems: (1) concurrent
execution of heterogeneous tasks on the same pilot; (2) support of all the
major HPC batch systems; (3) support of more than twelve methods to launch
tasks; and (4) a general purpose architecture. RP can execute single or multi
core tasks within a single compute node, or across multiple nodes. RP isolates
the execution of each tasks into a dedicated process, enabling concurrent
execution of heterogeneous tasks by design.

\section{Performance Measurement}
\label{sec:measure}

To understand the performance of \project it is imperative to
understand the nature of the computational campaign, its composite workflows,
their constituent workloads, desired performance and factors determining
scalability.
\subsection{Computational Characteristics}

\begin{figure}[h]
  \centering
  \includegraphics[width=.8\columnwidth,trim=6mm 0 0 0,clip]{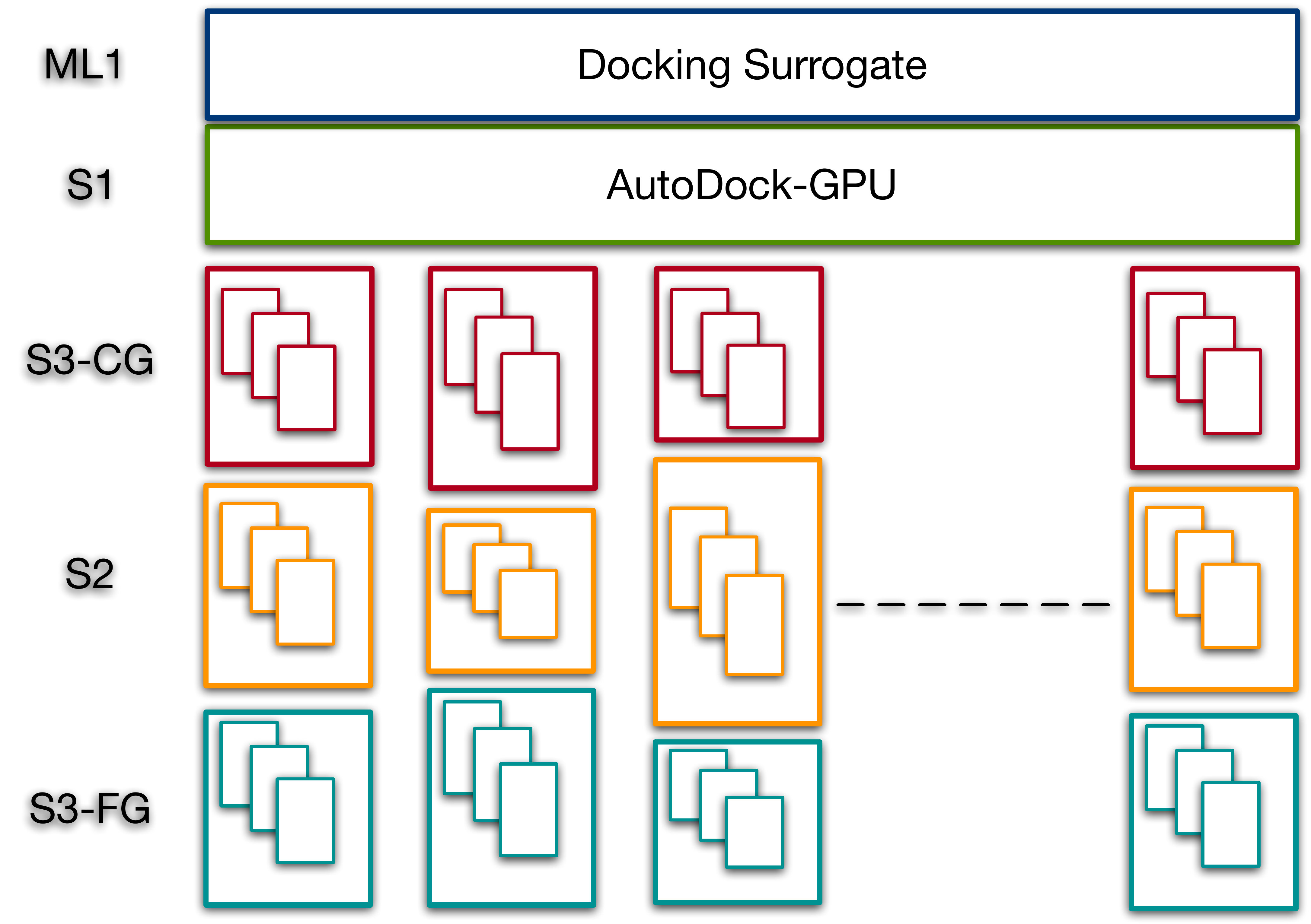}
  
    \vspace{-3mm}
    
  \caption{Programming and execution view: Each stage of the
  (S3-CG)-(S2)-(S3-FG) pipeline comprises multiple heterogeneous tasks; each
  stage executes for varying durations.}
  \label{fig:programming-framework}
\end{figure}

Fig.~\ref{fig:programming-framework} provides an overview of how
the
\project campaign is constructed and executed. It comprises four
distinct computational workflows: a machine learning surrogate
(ML1), docking (S1), binding free energy calculations (S3), and latent space
representation and steered advanced sampling via MD simulations (S2). Each is a distinct workflow with well-defined inputs and outputs,
multiple executables with defined dependencies, and termination criteria, able to
produce stand-alone scientifically meaningful end-results.
Each workflow represents the
expertise and unique scientific and methodological contribution from a
different team.

We codify \project workflows as a five-stage EnTK pipeline using a
general-purpose language (Python) and application-specific constructs from the
PST (Pipeline, Stage, Task) programming model. These abstraction simplify
creating and executing ensemble applications with complex coordination and
communication requirements. Pipelines can contain different workloads, e.g.,
distinct instances of S* for a given LPC, but also possibly multiple instances
of a given S* for many LPC concurrently. Autodock-GPU is executed as a single
task running on several thousand nodes, as is the docking surrogate, which is
a relatively short duration task. The remaining three stages are workflows which
are expressed as pipelines, comprised of differing stages and varying duration
and number of tasks concurrently executing. The horizontal length of a box is
proportional to the number of nodes utilized by a stage / computation, and the
vertical length of boxes represent the temporal duration; boxes are not drawn
to scale (see Table).

\subsubsection{ML1: Deep Learning Docking Emulator}
This step is a docking emulator which serves as a pre-selection tool for
docking calculations performed in step S1. The goal is to reduce the search
space from about 126M ligands down to a manageable amount for the docking
calculations.\ The emulator is based on a resnet-50~\cite{he2015deep} deep
neural network: it transforms image representations of ligand molecules into a
docking score. To convert the ligand SMILES strings into images we employed
the mol2D drawing submodule from rdKit~\cite{rdkit-mol-draw-2020}. The target
scores are  binding energies which are mapped into the interval [0, 1],
with higher scores representing lower binding energies and thus higher docking
probabilities. The main computational motifs are dense linear algebra,
convolutions and elementwise operations on 4D tensors. The network is
implemented in PyTorch and pre-trained on 500,000 randomly selected samples from the OZD
ligand dataset across each receptor (for our purposes, each PDB entry corresponds to a separate receptor, providing access to an ensemble of docking simulations). For deployment, we compiled the model
using NVIDIA TensorRT v7.2~\cite{trt-blog-2020} with cuDNN
v8.0~\cite{chetlur2014cudnn} employing the \texttt{torch2trt} helper
tool~\cite{torch2trt-2020}. As base precision we chose half precision (FP16),
so that we can use the Tensor Cores on Summit's V100 GPUs. 

Inference
workloads are notoriously IO bound, and thus we employ various optimizations to
improve throughput.
We start with the ULT911 dataset~\cite{ult911}, which is supplied as a collection of
12,648 files with 10,000 ligands, each in Python pickle format. We first
used gzip to compress each file, achieving an average compression factor of 14.2. We
use MPI to distribute the individual files evenly across a large number of
GPUs and bind one rank to each GPU. While we perform the model scaffolding
phase, i.e. creating the computational graph and loading the weights from the
pre-trained model file, each rank stages its assigned shard of the data from
GPFS into node-local NVME. During the inference process, each rank utilizes
multiple data loader processes where each is employing 2 prefetching threads:
the first one loads compressed files from NVME into DRAM and decompresses them
on the fly while the second iterates through the uncompressed data in memory,
extracts the image and metadata information and feeds them to the neural
network. The whole logic is implemented using the thread-safe python queue
module. We further use careful exception handling to make the setup resilient
against sporadic IO errors.\ After inference is done, the resulting lists of
docking scores and metadata information such as ligand id and SMILES string
are gathered and concatenated on rank 0 and written into a CSV file which is
forwarded to step S1.

\subsubsection{S1: Physics-based Ensemble Docking}

\begin{figure}
  \begin{center}
    \includegraphics[width=\columnwidth,trim=1mm 0 4mm 1.5mm,clip]{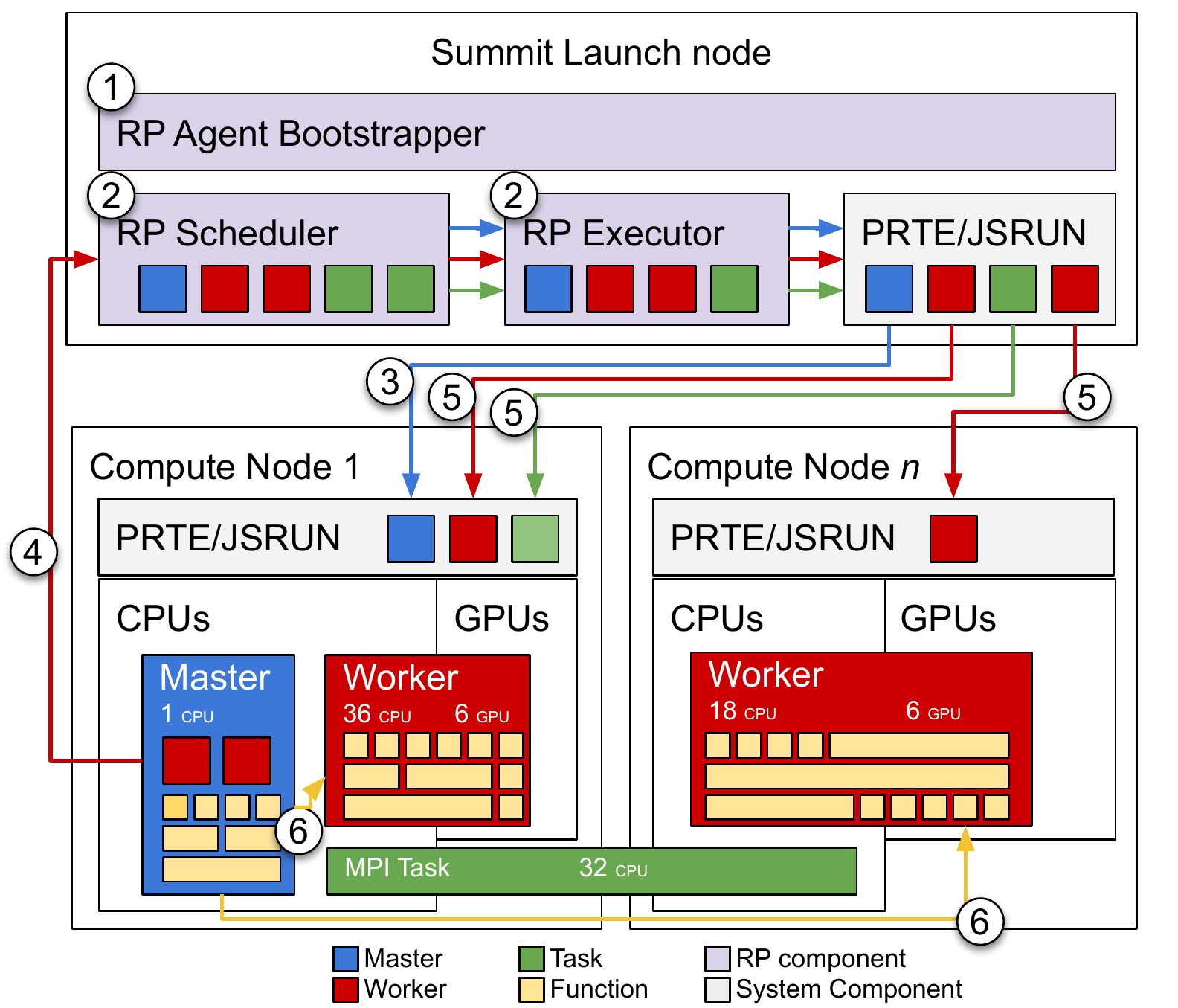}
  \end{center}
  
    \vspace{-3mm}
    
  \caption{RAPTOR Execution Framework: One of the two execution frameworks
  used to support heterogeneous tasks and dynamic workloads on
  Summit.}\label{fig:raptor}
\end{figure}

To support the scaling requirements of S1, we implemented a Master/Worker
overlay on top of the pilot-job abstraction. Fig.~\ref{fig:raptor} illustrates
the RAdical-Pilot Task OveRlay
(RAPTOR) master/worker implementation on Summit. Once RAPTOR has acquired its resources by submitting a job
to Summit's batch system, it bootstraps its Agent (Fig.~\ref{fig:raptor}-1)
and then launches a task scheduler and a task executor
(Fig.~\ref{fig:raptor}-2). Scheduler and Executor launch one or more masters
on one or more compute nodes (Fig.~\ref{fig:raptor}-3). Once running, a master
schedules one or more workers on RP Scheduler (Fig.~\ref{fig:raptor}-4). Those
workers are then launched on one or more compute nodes by RP Executor
(Fig.~\ref{fig:raptor}-5). Finally, the master schedules function calls on the
available workers for execution (Fig.~\ref{fig:raptor}-6), load-balancing
across workers so to obtain maximal resource utilization.

The duration of the docking computation varies significantly between
individual receptors.  The long tail poses a challenge to load balancing; the
relatively short docking times pose a challenge to scalability. Load balancing
is addressed by iterating through the list of compounds in a round-robin
fashion, and by dynamic load distribution which depends on the load of the
individual workers. Further, balancing is achieved by: (i) tasks are
communicated in bulks as to limit the communication load and frequency; (ii)
multiple master processes are used to limit the number of workers served by
each master, avoiding respective bottlenecks; (iii) multiple concurrent pilots
are used to isolate the docking computation of individual compounds within
each pilot allocation.  The combination of these approaches results in a near
linear scaling up to several thousand nodes, while maintaining high
utilization for large numbers of concurrently used nodes.

\subsubsection{S2 and S3: Advanced Sampling and Binding Free Energy}

We implement S2 and S3 as iterative pipelines that comprise heterogeneous
stages, with each stage supporting the parallel execution of tasks. In S2, the
pipeline starts with MD simulations that are run concurrently; it completes a
single iteration by passing through deep learning stages for AAE model
training and the outlier detection. In a single iteration, tasks are scheduled
across single GPU, multiple GPUs, and CPU-GPU tasks. For instance, the MD
stage uses a single GPU per simulation (OpenMM), the data aggregation stage
uses CPUs only, the ML training stage uses six GPUs per model, and the outlier
detection stage uses a mixture of CPUs and GPUs. We also employ data
parallelism for model training using PyTorch distributed data parallel module.

Similarly, S3 involves two stages of equilibration and one stage of
simulation; each stage runs an ensemble of from six (S3-CG) to 24 (S3-FG) OpenMM tasks. We also employ NAMD-based TIES in
conjunction with ESMACS; this requires placing distinct simulations to GPU
(OpenMM) and CPU (NAMD) concurrently for the optimal resource utilization on Summit.

The architecture of RAPTOR (Fig.~\ref{fig:raptor}) differs from that of the classic
RADICAL-Pilot used for S2 and S3 on Summit~\cite{turilli2019characterizing}.
The need for two task execution frameworks arises primarily from
the dynamism and heterogeneity of workloads. For example, six orders of
magnitude difference in the temporal duration (Tab.~\ref{tab:cost}) of tasks
requires that the Master/Worker overlay sustain a throughput of up to
50M docking hits per hour on $\sim$1000 nodes on Summit.

\section{Performance Results}
\label{sec:results}

We report on both scientific and computational performance.


\subsection{Scientific Performance}

Available compound libraries are large, with ZINC providing over 230 million purchasable compounds~\cite{Sterling_2015} in ready-to-dock, 3D formats, and MCULE having 100 million purchasable
compounds in similar formats~\cite{ult911}. Hence there is a need to obtain an appropriate
sampling of the compound libraries based on the diversity of the compounds as
well as their availability for both docking calculations (to generate the
training data) and inference runs (ML1 results). We selected a subset of 6.5
million compounds from the ZINC library along with the Enamine diversity set~\cite{enadiversity}
and DrugBank~\cite{Wishart_2017} compounds to develop our training library (OZD library,
hereafter). We also chose a similar subset of 6.5 million compounds from the
MCULE library (ORD library, hereafter) for the purposes of testing if  ML1 can
indeed be used for \emph{transferring} knowledge learned from one library to
another. These libraries pay attention only to the diversity of the compounds
selected, and are independently selected, although between the two libraries
we observed an overlap of approximately 1.5 million compounds.

\begin{figure}
    \centering
    \includegraphics[width=\columnwidth,trim=10mm 3mm 23mm 10mm, clip]{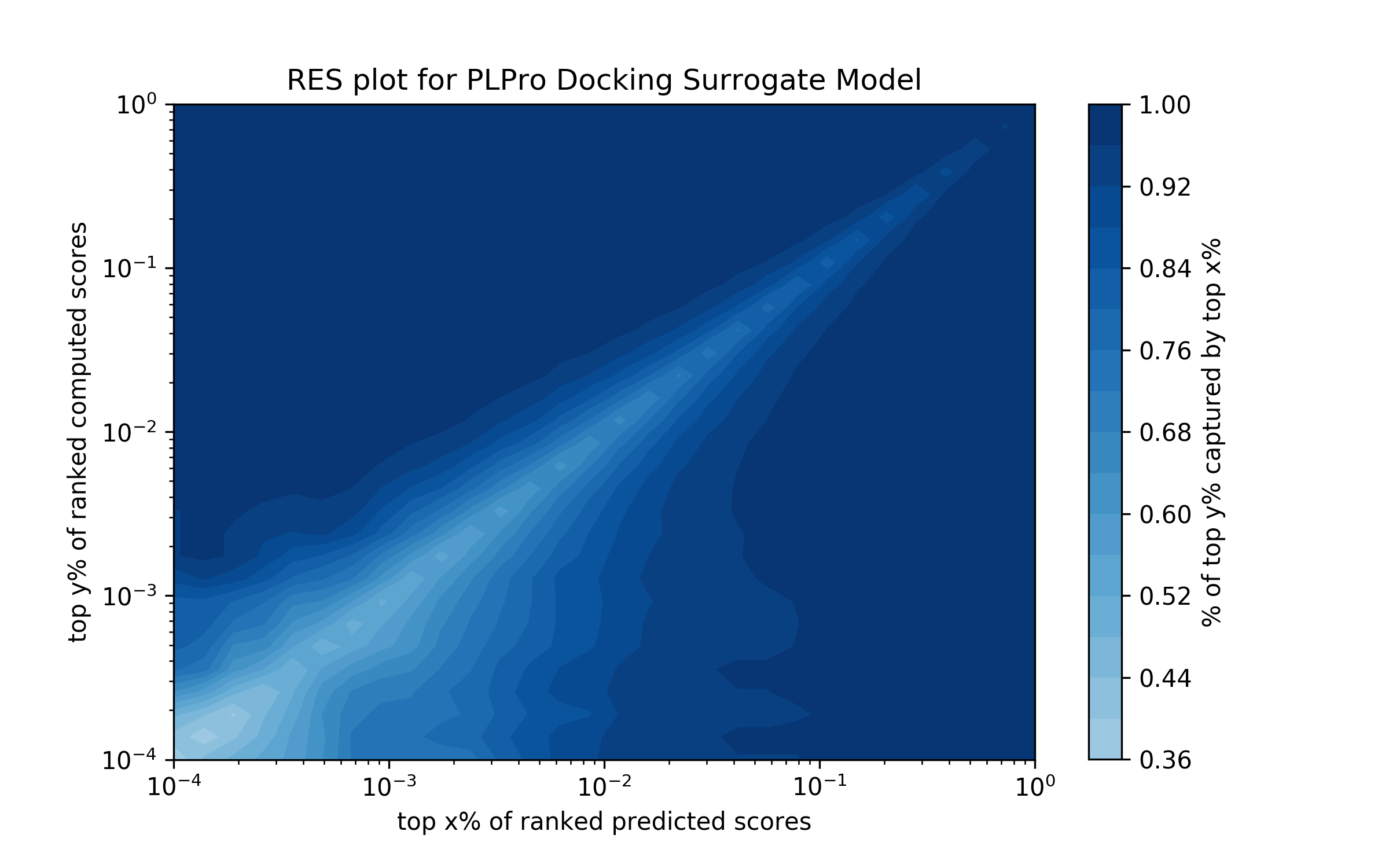}
    
      \vspace{-3mm}
      
    \caption{RES profile for PLPro docking runs. As explained in the main text, RES provides a summary estimate of how many top scoring compounds can be covered given some target number ($\delta$) of molecules to be ranked.}
    \label{fig:adrp_res}
\end{figure}
\subsubsection{ML1 results} We trained our ML1 models on docking runs
(generated offline) for the four main target SARS-CoV-2 proteins, namely 3C like
protease (3CLPro), papain-like protease (PLPro), 
ADP-Ribose-1"-Monophosphatase (ADRP), and non-structural protein 15 (NSP15). These proteins all represent important drug targets against
SARS-CoV-2 virus. Here we present only a vignette of results from the
PLPro target and specifically from the receptor derived from the PDB-ID 6W9C. The RES plot from Fig.~\ref{fig:adrp_res} indicates that the extraction of the top scoring
molecules from the OZD library can be thought of as a selection process of
$\delta$ molecules, to the subsequent stages. Given a specific budget of
$\delta$ molecules to pass along, we can imagine a vertical line along the
$x$-axis of Fig.~\ref{fig:adrp_res} at the point $\delta/u$
representing the budget, where any point to the right of
that line represents an unattainable number of compounds. One can also imagine
a constraint through $y=x$, as points above this line represent situations
where a wider range of the top distribution may prove too expensive, although reasonable for some tasks, uHTS screening is in pursuit of
ultra high ranking compounds.

Given these two constraints, one can see that as $\delta$ increases, so to does the accuracy
of capturing some desired threshold of the top distribution. If
downstream tasks allot $\delta=u10^{-3}$ compounds, then the plot indicates that we
will capture 50\% of the top ranking $u10^{-4}$ compounds, or around 40\%
of the top ranking $u10^{-3}$ compounds. In concrete terms, for this library,
the ML model here correctly identifies 500 of the top 1000 scoring compounds
from the docking study, or about 4000 of the top 10,000 compounds. However,
not all top-ranking compounds are correlated with obtaining high binding
affinity to PLPro. The RES plot also provides a quantitative estimate of the
number of compounds we have to sample from lower ranking one so that we do not
inadvertently miss out other high affinity compounds. Hence we also select
about 15--20\% of compounds from the RES to the subsequent stages.

\begin{figure}[h]
    \centering
    \includegraphics[width=\columnwidth,trim=3mm 0mm 3mm 2mm,clip]{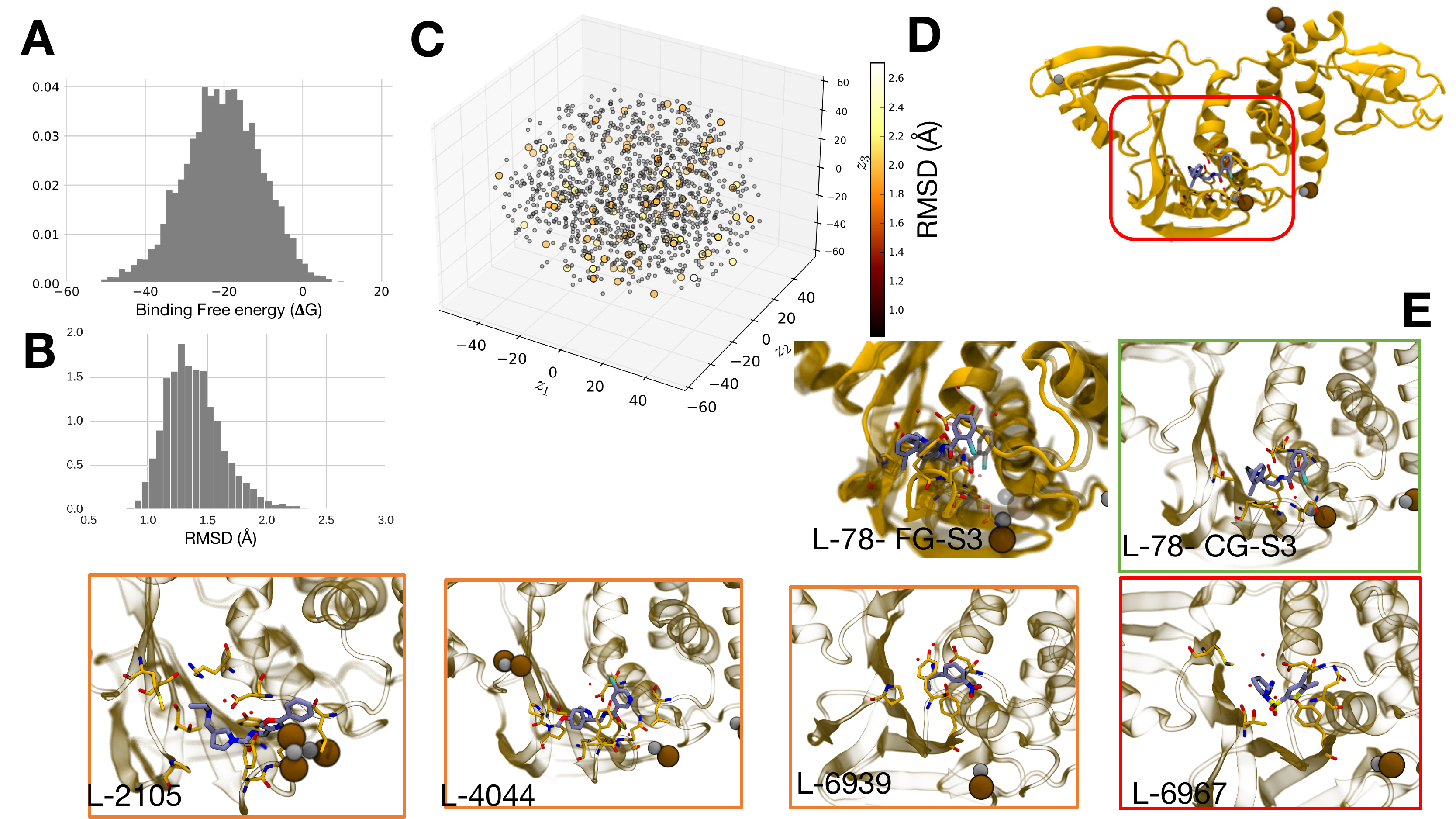}
    
      \vspace{-3mm}
      
    \caption{Preliminary results from \project on PLPro receptor (PDB ID: 6W9C). (A) Summary histogram of the distribution of binding free energies estimated using CG-ESMACS. (B) Summary of RMSD (\AA) determined from CG-ESMACS LPC ensembles show a rather tight distribution with a few LPCs that exhibit greater fluctuations. (C) Latent space representation from the 3D-AAE model depicting the outliers from  RMSD distributions (>1.9 $\AA$) and the rest as gray dots. The latent space also summarizes the extent of sampling from these simulations. (D) Structure of PLPro bound to one of the highly specific molecule (L78) in its active site. (E) A zoomed in version of the same compound (L78) showing close interactions with key residues in PLPro (green highlight). The panel on the left depicts how upon running FG-ESMACS we obtain tighter binding through the compound moving further into the binding site, forming strong hydrophobic interactions and hydrogen bonds. Other compounds that did not perform as well in the FG-ESMACS (see \ref{fig:ESMACS}) are shown for comparison; although the bound structures exhibit similar interaction patterns, none of them stabilize further in the binding site, leading to a reduction in their binding free energy estimates.}
    \label{fig:my_label}
\end{figure}

\subsubsection{S3: CG-ESMACS}
For each target of the four target proteins mentioned above, multiple crystal
structures were used to perform docking and a separate list of top 10,000
compounds based on their docking scores was generated at the ML1 stage.
Therefore, depending on the number of crystal structures used for each target,
there were collectively 20,000-40,000 compounds available for performing
binding affinity predictions using CG-ESMACS. At this stage of our pipeline,
we chose 10,000 compounds for each target by picking out the structurally most
diverse compounds from all compounds available. This was done for two reasons:
(i) based on the docking scores, all the available compounds were stable
poses, and (ii) allowing for maximum possible coverage of the chemical space
allowing for better and quicker identification of its relevant regions to
focus on in next iterations.

We performed CG-ESMACS to get binding affinities for all these compounds
chosen, amounting to a total of 40,000 S3-CG calculations thus far. Fig.~\ref{fig:my_label}A shows a
probability distribution of the 10,000 binding affinities for PLPro. The values typically lie
between -60 to +20 kcal/mol for all proteins. The resultant trajectories and
binding affinity values from this stage were used as input for S2 to identify
potentially useful conformations that were fed into S3-FG.

\subsubsection{Using S2 to seed S3-FG} For PLPro, about 5000 compounds were
chosen based on the structural diversity criterion for PDBID 6w9c. The
trajectories corresponding to these 5000 compounds generated by S3-CG were
used to build a combined dataset of 100,978 examples. The point
cloud data, representing the coordinates of the 309 backbone C$^\alpha$ atoms
of the protein, was randomly split into training (80\%) and validation input
(20\%) and was used to train the 3D-AAE model for 100 epochs using a batch
size of 64. The data was projected onto a latent space of 64 dimensions
constrained by a Gaussian prior distribution with a standard deviation of 0.2.
The loss optimization was performed with the Root Mean Square Propagation
(RMSprop) optimizer, a gradient descent algorithm for mini-batch learning,
using a learning rate of 0.00001.

We also added hyperparameters to scale individual components of the loss. The
reconstruction loss was scaled by 0.5 and the gradient penalty was scaled by a
factor of 10. We trained the model using several combinations of
hyperparameters, mainly varying learning rate, batch size and latent
dimension. The embedding learned from the 3D-AAE model summarizes a latent
space that is similar to variational autoencoders, except that 3D-AAEs tend to
be more robust to outliers within the simulation data. The embeddings learned
from the simulations allow us to cluster the conformations (in an unsupervised
manner) based on their similarity in overall structure, which can be typically
measured using quantities such as root-mean squared deviations (RMSD). The
5,000 ligands were further analyzed and 5 structures with the lowest free
energy (L6967, L2105, L78, L6939, L4044) were selected for generating
embeddings for 1200 examples, using the hyperparameters learned from 3D-AAE
performed on the full set of 5,000 ligands. For visualizing and assessing the
quality of the model in terms latent space structure, we computed
t-stochatstic neighborhood embedding (t-SNE)~\cite{maaten2008visualizing} on the
 embeddings from the validation set. The validation data was
painted with grey while the test data was painted with the root mean squared
deviation (RMSD) of each structure to the starting conformation (Fig.~\ref{fig:my_label}B-C).

\subsubsection{S3: FG-ESMACS}

\begin{figure}
    \centering
    \includegraphics[width=0.9\columnwidth,trim=5mm 12mm 30mm 30mm,clip]{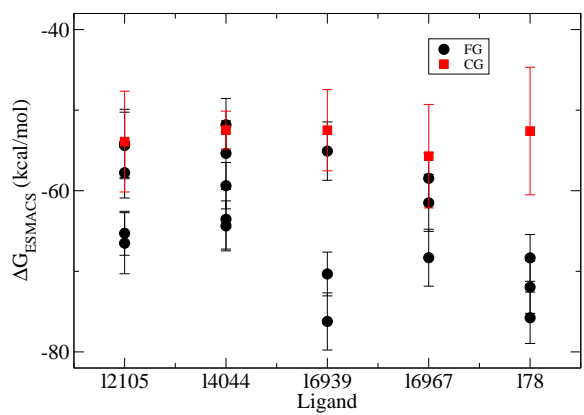}
    
      \vspace{-3mm}
      
    \caption{Comparison of S3-CG and S3-FG results for the five best binders for
    PLPro (PDBID: 6w9c) based on CG-ESMACS results. S2 selected five outlier
    conformations for each binder and performed FG-ESMACS on them. The
    provisional results confirm improved binding for the selected conformations
    in all five compounds, as FG energies are lower than CG.}
    \label{fig:ESMACS}
    \vspace{-0.1in}
 \end{figure}

The large amount of data generated by S3-CG was analysed at S2 and based on
that potentially good conformations were identified for the compounds with
large negative binding affinities from CG-ESMACS. This process led us to
filter out five outlier conformations each for the top five compounds based on S3-CG
results. We used these 25 conformations to perform the costlier
FG-ESMACS calculations to demonstrate the capability of our pipeline to
confidently identify favourable interactions between protein and ligands. This
helps us mark favourable regions in the chemical space deserving more
attention, which in turn trains our ML model to generate and/or predict better
compounds in the next iteration. Fig.~\ref{fig:ESMACS} displays a comparison
of the provisional results from FG-ESMACS with those from CG-ESMACS. It is manifest
that FG-ESMACS predicts much lower binding affinities than those predicted by
CG-ESMACS (Fig.~\ref{fig:my_label}D-E). 
The force-field used in both cases was the same; only the starting
structures varied between them. This implies that the outliers filtered out by
S2 indeed captured some favourable interactions and successfully identified
good conformations out of the large number of conformations generated by
S3-CG. This is an excellent demonstration of the novel capability of \project
to quickly sample the relevant chemical space and hence accelerate the process
of drug discovery. With every iteration, such refinements improve the
compounds generated by our ML model manifold: we expect to design potential
inhibitors for the target protein in much less time than the standard drug
discovery approaches.

\subsection{Computational Performance}

\begin{figure}[h]
  \centering
  \includegraphics[width=\linewidth]{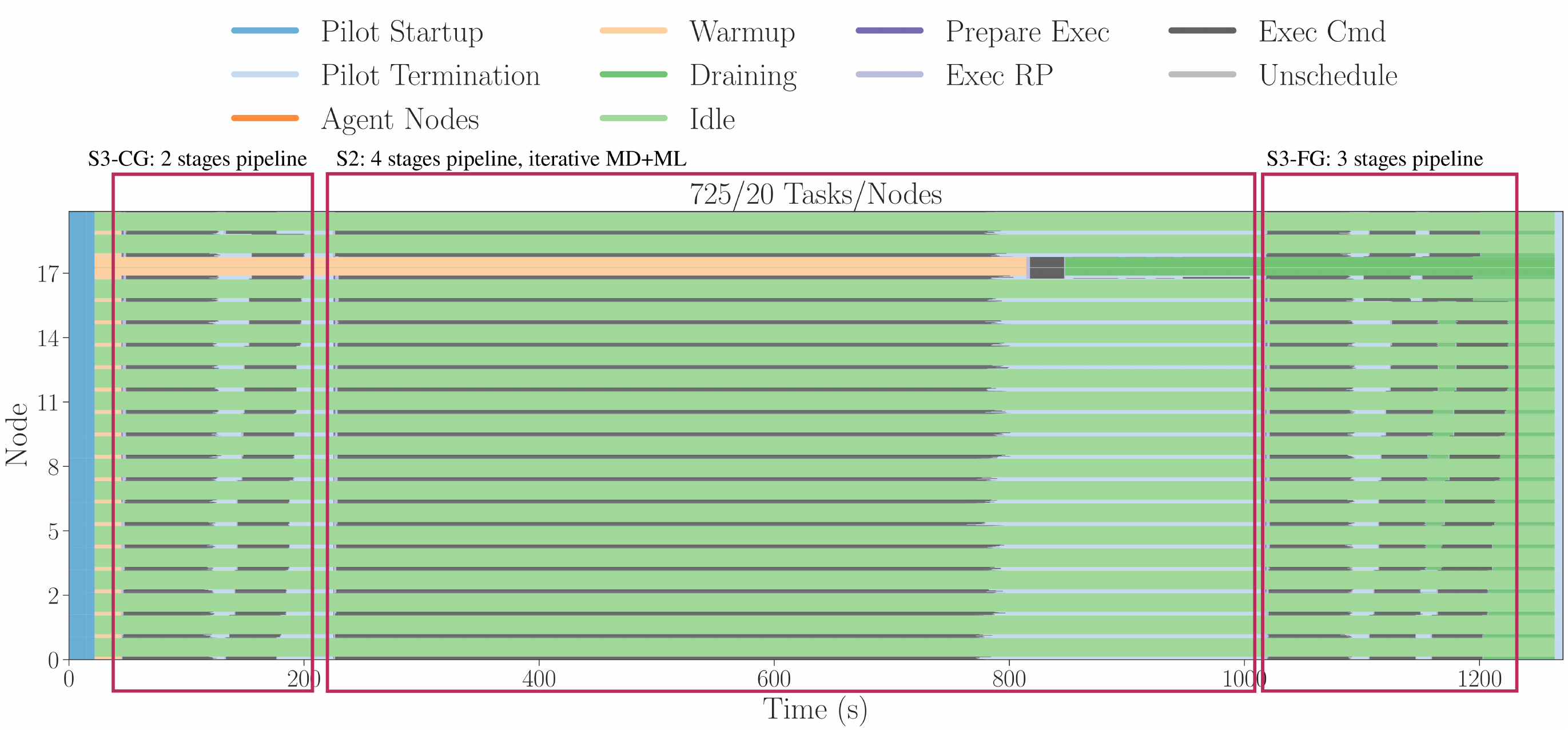}
  
  \vspace{-3mm}

  \caption{A time-series of node utilization. The Fig.~depicts the integrated
  execution of three GPU-intensive workflows (S3-CG)-(S2)-(S3-FG). S3-CG, S2 and
  S3-FG are heterogeneous and multi-stage workflows themselves.}
  \label{fig:s3-s2-s3-ru}
\end{figure}

Fig.~\ref{fig:s3-s2-s3-ru} shows an example of how independent pipelines can be
integrated into a single workflow. Each pipeline is comprised of stages, each
with an arbitrary number of tasks. Tasks have heterogeneous execution time and
computational requirements. Stages can execute concurrently or sequentially,
depending on available resources and task, stage and pipeline interdependencies.
In the depicted integration, single-GPU tasks execute alongside MPI GPU and few
CPU tasks, in distinct and customized execution environments. Note that the
overheads (light-colored vertical areas of the plots) are invariant to scale,
i.e., they do not depend on the number of concurrent tasks executed or on the
length of those tasks.

We measure flops (floating point operations, not rates) per
work unit for the most relevant components of each stage. We define a \emph{work
unit} to be a representative code section such as an MD time integration
step for MD-based or a data sample for DL-based
applications. Thus we can compute the aggregate invested flops by
scaling the measured flop counts to the respective work set sizes used in the
actual runs.

\begin{table}[h!]
 \caption{Throughput and performance measured as peak flop per second (mixed
precision, measured over short but time interval) per Summit node (6 NVIDIA
V100 GPU).}
\label{tab:performance}
\vspace{-0.1in}
\begin{tabular}{c|rrr}
\hline
Comp. & \#GPUs & Tflop/s & Throughput \\
\hline
\hline
ML1 & 1536 & 753.9 & 319674 ligands/s \\
S1  & 6000 & 112.5 & 14252 ligands/s \\
S3--CG & 6000 & 277.9 &  2000 ligand/s \\
S3--FG & 6000 & 732.4 & 200 ligand/s \\
\hline
\end{tabular}

\vspace{-0.2in}
\end{table}

We always normalize the measurements to a single Summit node for the same
reason. As all of our applications are perfectly load balanced with respect
to a Summit node (mostly even with respect to individual GPUs within that node),
this procedure yields a representative flop count. We use the
methodology of Yang et al.~\cite{yang2020hierarchical} and the NVIDIA NSight
Compute 2020 GPU profiling tool to measure flops for all precisions and sum them
to obtain a mixed precision flop count. When possible, we use start/stop
profiler hooks to filter out the representative work units. In order to obtain
the flop rate, we divide the aggregated flops for each EnTK task by the time it
takes to complete that respective task, including pre- and post-processing
overhead. Note that we do not account for any CPU flops invested in this
calculation as we expect that number to be small. We discuss the specifics for
each component:

\vspace{0.5ex}
\noindent
\emph{ML1}: We count flops as described above for 10 steps
at batch size 256. From that, we derive a flop count per batch per GPU.

\vspace{0.5ex}
\noindent
\emph{S1}: We count flops for a five-ligand AutoDock-GPU run on one GPU
to derive flops for a single ligand. We chose this ligand complexity 
to represent the majority of the ligands processed in the run.

\vspace{0.5ex}
\noindent
\emph{S2}: This stage has multiple steps, but we only account for the
autoencoder training and the MD performed in this stage. For the former, we
measure the flops per batch for a batch size of 32 for training and validation
separately and weight them proportionally by their relative number of batches.
After each training epoch, a validation is performed and the train/validation
dataset split is 80\%/20\%. This can be translated into a overall flop count for
the full autoencoder stage. For the MD part, we profile 20 steps of OpenMM and
compute a complexity per step.

\vspace{0.5ex}
\noindent
\emph{S3-CG/FG}: These two stages both have two
steps, a minimization and an MD step. We count the flops for 10 iterations of
the minimization algorithm and for 20 steps of the MD run to derive a flop count
per minimization and MD step. Since the algorithmic complexity differs between
CG and FG, we profile those separately.

\section{Implications}

Multiscale biophysics-based computational lead discovery is an 
important strategy for drug development and while it can be considerably
faster than experimental screening it has been until now too slow to explore
libraries of the scale of billions of molecules even on the fastest machines.

The work reported here not only addresses this performance issue, by demonstrating a path towards an overall improvement of throughput of computational drug discovery of order 1000x by integrating machine learning components with the physics based components, but it also addresses other important aspects of improved workflows for computational lead discovery, namely the generalization and integration of feedback between the physics models and the ML models in a tightly coupled workflow.

By introducing ML modules paired with and trained from the physics modules output, over time the ML component models improve such that the overall workflow becomes tuned to the specific target problem.  It is also likely that it will be possible to transfer these learnings from one problem to the next.  The generalizability of this approach is under active investigation.

The feedback is not only local to each stage of the workflow but end-to-end, so this work also represents the beginning of an autonomous drug development system when coupled to automated experimental screening and eventually to drug synthesis capability.  

The work reported here has played a central role in the DOE effort to use
computational molecular design approaches to develop medical therapeutics for
COVID-19.  DOE established in April 2020 the National Virtual
Biotechnology Laboratory (NVBL)~\cite{nvbl}, to organize the DOE national
laboratories into a series of projects---including therapeutics
development---aimed at addressing key challenges in responding to
COVID-19.  Methods and infrastructure reported in this paper are being used to
screen over 4.2 billion molecules~\cite{babuji2020targeting} against over a dozen drug targets in
SARS-CoV-2.  This work has already lead to the identification and experimental
validation of over 1000 compounds, resulting in over 40 hits that are
progressing to advanced testing.

To get a sense of the scale of operations using methods and infrastructure
reported here: in the past three months, we have used more than \textbf{2.5M
node-hours} across diverse HPC platforms, including TACC's
Frontera, Livermore Computing's Lassen, ANL's Theta (and associated A100
nodes), LRZ's SuperMUC-NG, and ORNL Summit to obtain scientific
results. We have executed individual parts of the 
campaign on suitable platforms, e.g., we sustained 40M docking hits
per hour over 24 hours on 4000 nodes on Frontera. We 
performed thousands of AI-driven MD simulations across GPU-based systems: Lassen, Summit and Theta's A100
nodes, as well as a few tens of thousands of CG-ESMACS runs on Summit, ARCHER (EPCC, UK), Monsoon2 (Metoffice, UK) and SuperMUC-NG.

In doing so, \project has screened $\sim$10$^{11}$
ligands. We have performed up to $5\times 10^{7}$ docking-hits per hour
using OpenEye and Autodock-GPU, and sustained this throughput on $\sim$4000
nodes. Individual workflow components deliver 100$\times$ to 1000$\times$ improvement
over traditional methods.
\project has computed binding free energies on 10$^{4}$ LPCs
concurrently.

While much work remains to be done, we have demonstrated some important
milestones towards the ultimate goal. These include: orders of magnitude end-to-end
performance improvement of computational methods for drug discovery, making it
feasible and routine to search giga-scale libraries of compounds across
collections of drug targets; integration of physics based modeling with AI
methods into multiscale workflow at the largest-scale possible providing a
pathway for exascale drug discovery, and developed a prototype infrastructure
that can be adapted to a broad range of near autonomous drug development
scenarios by means of additional modules and models that fill out the
drug discovery pipeline.

We have developed this campaign with multiple target problems in mind. A major
influence is our experience in working on the DOE/NCI JDAS4C Pilot1 effort to
advance Cancer drug development through AI~\cite{pilot1} and the related ECP
CANDLE project~\cite{candle}.  In the Pilot1 activity we are building Cancer
drug response models that predict the response of tumors to drugs or drug
combinations. Such models could be coupled to the the workflow described in
this paper to add additional feedback on predicted efficacy of a target
molecule.  Work in this area is ongoing to build a bridge between tissue level
efficacy models such as the ones developed in Pilot1 and CANDLE and drug
target oriented models reported here.

We envision ultimately a series of models to be part of this extended workflow
that predict not only efficacy, but also other important properties of
candidate compounds such as drug metabolism, pharmacokinetics, absorption,
toxicity, distribution, and excretion that traditionally would be assessed via
experimental methods.  Building a comprehensive suite of ML-based predictors
is precisely the goal of the ATOM consortium, to which this project is a
contributor.  ATOM aims to dramatically accelerate drug development by
building and training ML models that systematically replace routine experimental
screening in early stages of drug development, with the goal of reducing the time
from drug target to clinical trial from 5--6 years to one year.

\vspace{1ex}

\small{

\noindent {\bf Acknowledgements:} Research was supported by the DOE Office of
Science through the National Virtual Biotechnology Laboratory, a consortium of
DOE national laboratories focused on response to COVID-19, with funding
provided by the Coronavirus CARES Act. This research was supported as part of
the CANDLE project by the Exascale Computing Project (17-SC-20-SC), a
collaborative effort of the U.S. Department of Energy Office of Science and
the National Nuclear Security Administration. This work has been supported in
part by the Joint Design of Advanced Computing Solutions for Cancer (JDACS4C)
program established by the U.S. Department of Energy (DOE) and the National
Cancer Institute (NCI) of the National Institutes of Health. We are grateful
for funding for the UK MRC Medical Bioinformatics project (grant no.
MR/L016311/1), the UK Consortium on Mesoscale Engineering Sciences (UKCOMES
grant no. EP/L00030X/1) and the European Commission for the EU H2020
CompBioMed2 Centre of Excellence (grant no. 823712), as well as financial
support from the UCL Provost.  Access to SuperMUC-NG, at the Leibniz
Supercomputing Centre in Garching, was made possible by a special COVID-19
allocation award from the Gauss Centre for Supercomputing in Germany. Anda
Trifan acknowledges support from the United States Department of Energy
through the Computational Sciences Graduate Fellowship (DOE CSGF) under grant
number: DE-SC0019323. We acknowledge amazing support from OLCF\textemdash{}
Don Maxwell, Bronson Messier and Sean Wilkinson. We also wish to thank Dan
Stanzione and Jon Cazes at Texas Advanced Computing Center.}

\newpage 

\bibliographystyle{ACM-Reference-Format}
\bibliography{radical,references}

\end{document}